\documentclass[usenatbib]{mn2e}
\usepackage{psfig}

\newcommand{\ltaraw}{$\; \buildrel < \over \sim \;$}
\newcommand{\lta}{\lower.5ex\hbox{\ltaraw}}
\newcommand{\gtaraw}{$\; \buildrel > \over \sim \;$}
\newcommand{\gta}{\lower.5ex\hbox{\gtaraw}}

\loadboldmathitalic   \title[Cosmological   Radar  Ranging]{Cosmological
  Radar Ranging in  an Expanding Universe\footnotemark[1]} 
 \author[Lewis et al.]{Geraint F.
  Lewis$^{1}$, Matthew J. Francis$^1$,  Luke A. Barnes$^{2,1}$, Juliana Kwan$^1$ 
  \newauthor \& J. Berian James$^{3,1}$\\
  $^{1}$Institute of Astronomy, School of Physics, A28,
  University of Sydney, NSW 2006, Australia\\
  $^2$Institute of Astronomy, University of Cambridge, Madingley Rd, Cambridge, CB3 0HA, UK\\
  $^3$Institute for Astronomy, Royal Observatory, Edinburgh, EH9 3HJ, UK\\
} 
\date{\today}
\begin{document}
\maketitle
\label{firstpage}
\begin{abstract}
 While modern cosmology, founded in the language
of general  relativity, is  almost a century  old, the meaning  of the
{\it expansion of  space} is still being debated.   In this paper, the
question  of  radar ranging  in  an  expanding  universe is  examined,
focusing upon light  travel times during the ranging;  it has recently
been claimed  that this proves  that space physically expands. We  generalize the
problem into  considering the return  journey of an accelerating rocketeer, showing
that while this agrees with expectations  of special relativity for an empty
universe,  distinct  differences  occur  when  the  universe  contains
matter. We conclude that this  does not require the expansion of space
to  be  a physical  phenomenon,  rather  that  we cannot  neglect  the
influence of matter, seen through the laws of general relativity, when
considering motions on cosmic scales.
 \end{abstract}

\long\def\symbolfootnote[#1]#2{\begingroup%
  \def\thefootnote{\fnsymbol{footnote}}\footnotetext[#1]{#2}\endgroup} 

\def\newblock{\hskip .11em plus .33em minus .07em}
\section{Introduction}     \label{intro}     \symbolfootnote[1]{Research
  undertaken  as part  of  the Commonwealth  Cosmology Initiative  (CCI:
  www.thecci.org),  an  international  collaboration  supported  by  the
  Australian Research Council}

The  question  ``Is  space  {\it really}  expanding?''   has  recently
(re)surfaced  in the  literature, with  varying views  on  whether the
expansion of space  is a phenomenon which can  be directly observed.
\citet{2004Obs...124..174W} entered the fray with a Newtonian analysis
of particles detached from the  Hubble flow, showing their motion does
not agree to the simple viewpoint of expanding space-time as a form of
force.   \citet{2006astro.ph.10590C}   also  considered  the  physical
implications  of   the  expansion   of  space,  suggesting   that  the
superluminal expansion  of distant objects, often touted  as the proof
of  expansion,  can be  removed  with  a  transformation to  conformal
coordinates, and hence cannot be  physical, although it has been shown
that       superluminal       expansion       does       in       fact
remain~\citep{2007MNRAS.381L..50L}.         \citet{2007PASA...24...95F}
assessed the situation  in detail, showing that the  view of expanding
cosmologies as expanding space is a valid interpretation as
long as the equations of relativity are used to guide common-sense.

Recently, \citet{abram}  considered radar ranging of  a distant galaxy
in expanding cosmologies and concluded  that the fact that the radar and
Hubble distance, from  $d=H_o v$, differ in all  but an empty universe,
that space must really expand\footnote{The title of
  their  paper  `Eppur  si  espande'  is a  slight  rewording  of  the
  mutterings of Galileo  after his trail for heresy  by the Vatican in
  1633. It seems to demonstrate  that these authors believe that space
  `really'      expands.}.       In      a      counter      argument,
\citet{2006astro.ph..1171C}  again  considers  radar ranging  in  open
cosmological models.  Instead of  examining distances, he focuses upon
the transit  time of light  in usual cosmological coordinates  and its
conformal  representation. With  this he  reveals that  in  the former
coordinates the paths are  asymmetrical in transit time, taking longer
on  the return journey,  whereas in  conformal coordinates,  the light
travel  times to and  from the  distant galaxy  are equal.   Hence, he
concludes that the expansion of space is a coordinate dependent effect
which can be made to  disappear with the correct coordinate transform,
and therefore the expansion of space is not a physical phenomenon.

\begin{figure*}
\centerline{ \psfig{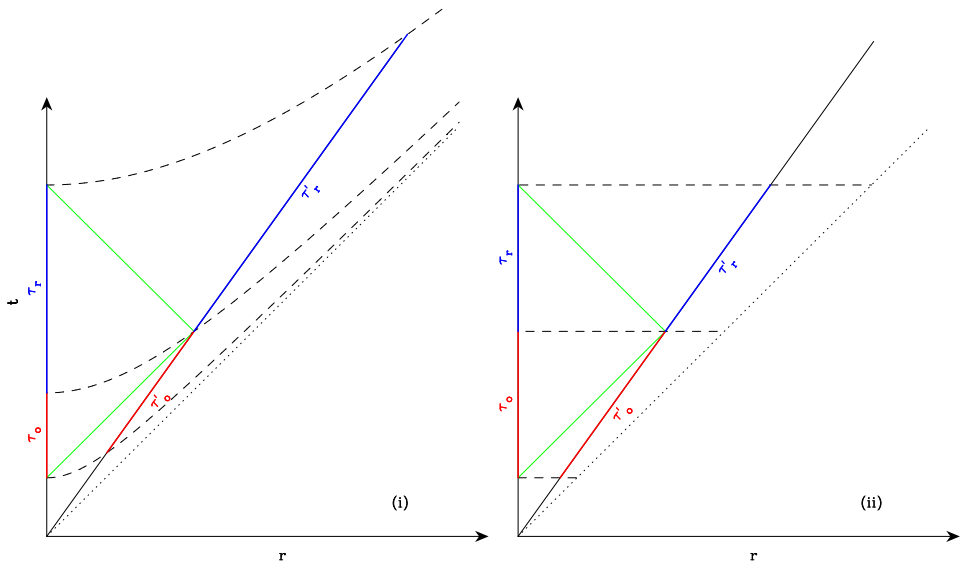}}
\caption[]{Radar ranging in a fully conformal representation of an open 
universe. In both, a light ray emitted from the origin is represented by 
a dotted path, while a comoving observer is represented 
by  the  solid sloping  line and  another  observer  sits at  the  origin
($r=0$).  The green line shows the null geodesic representing the laser
ranging  beam originating at  $r=0$, which is reflected  back by  the
distant comoving  observer. In the  left hand panel, the  dashed lines
represent  curves of constant  time in  the FLRW  metric, while  on the
right the  dashed lines represent intervals of constant  conformal time.
The red and  the blue paths represent the times measured by the
observers for the outward and returning rays respectively.
\label{fig1}}
\end{figure*}

In  this contribution  we examine  the recent  debate  on cosmological
radar ranging of  objects, clarifying some of the  issues discussed by
other  authors and demonstrate  that while  expanding space  remains a
useful concept, such experiments in no way require expanding space
as a physical effect.
The issue of   radar  ranging   in   Friedmann-Lemaitre-Robertson-Walker  
(FLRW) universes  is  addressed in Section~\ref{range}, conformal 
coordinates and generalized to include the motion of an accelerating  
observer in Section~\ref{clock}. A comparison of our results in 
light of previous studies is presented in Section~\ref{syn}, where 
we also offer our conclusions on the issue of expanding space.  

\section{Background}\label{range}

\subsection{FLRW \& Conformal Cosmology}\label{conform}
Modern cosmological models are described by the relativistic equations
for   a  homogeneous   and  isotropic   distribution  of   matter  and
energy. Many  elementary textbooks show how,  under these assumptions,
the spacetime  of the universe is  described by the  FLRW metric, whose
invariant interval is given by
\begin{equation}
ds^2 = dt'^2 - a^2(t')\left[ dx^2 +R_o^2 S^2_k(x/R_o) (d\theta^2
+ sin^2\theta\ d\phi^2) \right],
\label{flrw}
\end{equation}
where $S_k(x) =  \sin x, x, \sinh x$ for  spatial curvatures of $k=+1$
(closed),  $k=0$  (flat)  and  $k=-1$ (open)  respectively,  with  the
curvature given  by $R_o^{-2}$; note, $c=1$.   The scale factor,
$a(t')$, governs  the dynamics of  the expansion and is  dependent upon
the relative mix of matter and energy in the universe.

Conformal transformations preserve angles at a point and are important
in geometry. For the purposes of this study, the space-times of interest
are those which are `conformally flat', such that the metric of a curved
space time is related to that of the flat space-time of special
relativity via \begin{equation} g = \Omega({\bf x}) g_{flat}.
\end{equation} 
where $\Omega({\bf x})$ is an arbitrary function.
If we take a particular two-dimensional slice of the FLRW metric
above (Equation~\ref{flrw}) then it is easy to construct a conformally
flat transformation such that~\footnote{In fact, any two dimensional
subspace is conformally flat; see Appendix 11C of
\citet{2005gere.book.....H}} \begin{equation} ds^2 = a^2(\eta)( d\eta^2
- dx^2 ), \label{conf} \end{equation} defining the conformal time to be
related to the universal time through $dt' = a(\eta) d\eta$. Throughout
this paper, such a transformation will be referred to as a partial
conformal transformation; clearly, in the $(x,\eta)$ coordinates light
rays ($ds=0$) will trace out the undistorted light cones of special relativity.

Typically, conformal representations of FLRW cosmologies employ 
only the partial transform, although
this flattens the  radial part of
the metric,  and hence will  not in general  produce flat SR-like light
cones in fully 4D space-time. For a general  cosmology with spatial  curvature, a full
conformal transformation  is required  to make the  metric conformally
flat in all  four dimensions, as explored in  detail by \citet{is}, in
which the comoving radial coordinate  must be transformed as well. For
instance,  in an open  universe the  fully conformal  coordinates [see
\citet{2006astro.ph..1171C}  and  \citet{2007MNRAS.381L..50L} for  the
derivation and more details] are:
\begin{eqnarray}
r =Ae^\eta\sinh \chi \nonumber \\
t = Ae^\eta\cosh \chi
\label{fullconformal}
\end{eqnarray}
where $A$ is  a constant with $\chi = x/R_0$. We  have referred to the
fully  conformal coordinates  as $(r,t)$,  matching those  employed in
previous studies. It is now  expected that light rays  follow the
classic special  relativistic light cones in  these coordinates. Clearly,  since
the spatial  part of a flat $(k=0)$ FLRW metric is already  flat, the partial
and fully conformal transformations are equivalent. In this study, both the
partial and fully conformal transformations will be 
considered, to provide a comparison to previous studies and allow a correspondence 
with the flat space-time of special relativity as $\Omega_o\rightarrow 0$.

\section{Radar Ranging}
The principle of radar ranging is simple; to calculate the distance to
a distant object, fire a radar  pulse at it and time the
interval  $\Delta\tau$  until the  beam  returns; here $\tau$ is the 
proper time as measured by an observer sending out and receiving the 
radar beam.  From  this, it  is straightforward to define a radar distance
\begin{equation}
D_{rad} = \frac{\Delta\tau}{2}
\end{equation}
assuming  $c=1$. For  cosmological  cases,  it is  usual  to measure 
the comoving distance to a galaxy whereas the time is measured by 
an observer at the origin. 
Clearly this is a well defined experiment as we are asking about the 
time ticked off along a world line, an observable quantity.
Remember,  however, that \citet{2006astro.ph..1171C} 
  asked a  somewhat different  question, namely  how  much time
passes on the individual outward and return journeys.

\begin{figure*}
\centerline{ \psfig{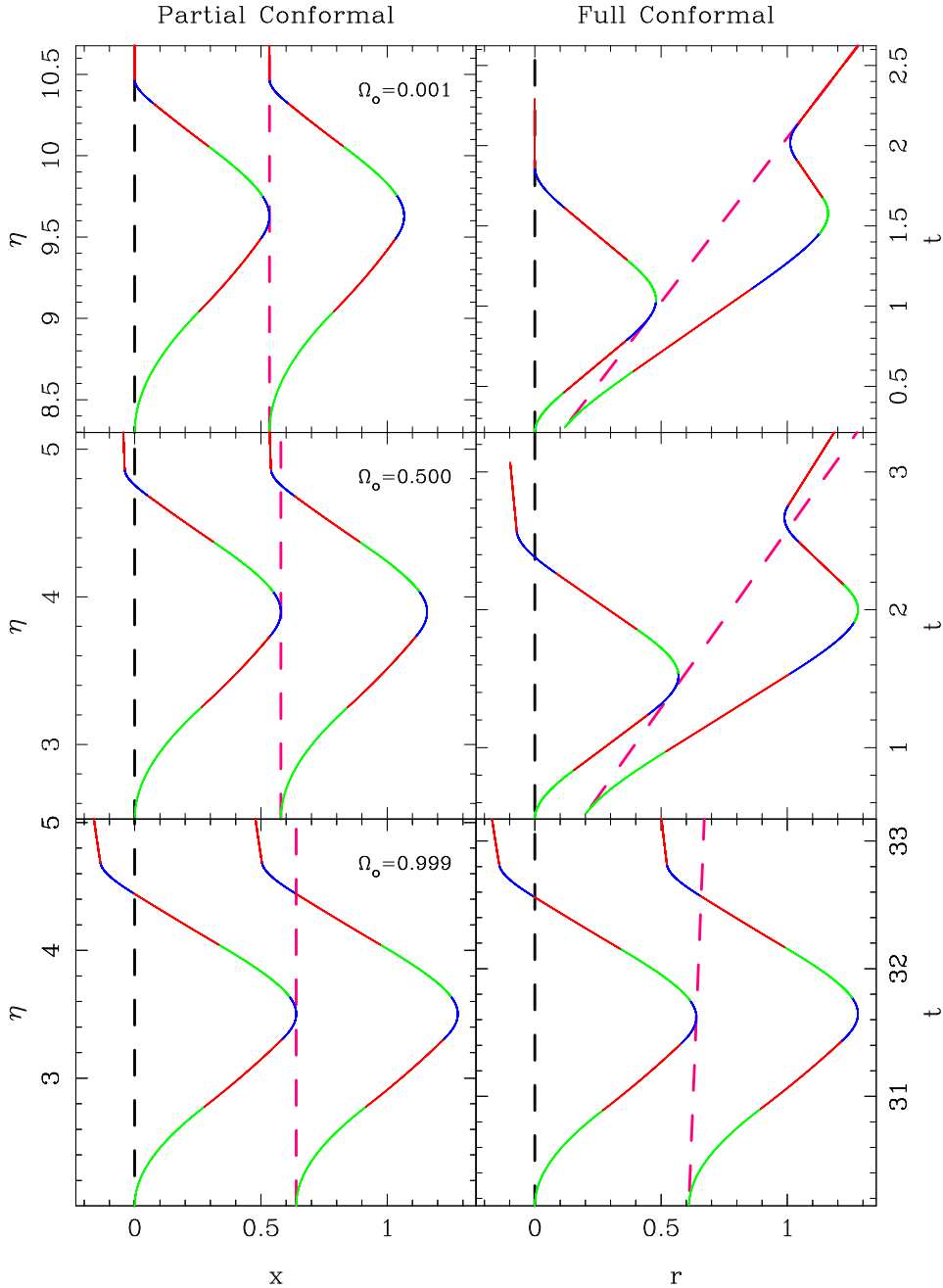}}
\caption[]{  Cosmological  radar  trips  for rocketeers  in  partially
conformal   (left   panels)  and   fully   conformal  (right   panels)
representations   of  three   open  universes   with   varying  matter
content.  For each  universe,  we  have considered  the  paths of  two
rocketeers, one  leaving from the  origin and another that  starts out
from an  arbitary comoving location.  The colour sections of  the path
describe the  state of the rocket  engine of the  rocketeer. The first
green region shows the  initial outwards acceleration, followed by the
red period corresponding to a coasting phase with no acceleration. The
rocket is then  swung around to accelerate towards  the origin and this
is reprented by  the blue and then green region.  The change from blue
to  green indicates  the midway  point in  this acceleration  phase as
measured  in the  rocketeers  time. Following  this  there is  another
coasting period,  shown in red and  finally the rocket  is swung around
again to accelerate  away from the origin and  this corresponds to the
final blue  region. The black  dashed lines are  the paths taken  by a
comoving  observers starting from  the origin,  while the  pink dashed
lines  are for  comoving  observers starting  from  the same  comoving
location as  the second rocketeer.  For the fully conformal  cases, we
have used  $A = 5.3\times10^{-5},  8.3\times10^{-2}$ and
28.3 for $\Omega_o = 0.001, 0.500$ and 0.999 respectively.
\label{fig2}}
\end{figure*}

Figure~\ref{fig1} presents  the radar  ranging experiment for  an open
universe  in fully conformal coordinates;  here a  radar pulse  (green line)
leaves  the origin,  and is  reflected back  from a  distant, comoving
(constant spatial FLRW coordinates) object,  later being received back 
at the origin.  Note that in  this conformal picture, comoving observers 
move along paths which originate at the origin and move along  a line of
constant slope given by
\begin{equation}
\frac{dr}{dt} = \tanh( \chi ),
\end{equation}
where $\chi$ is the  comoving coordinate of the fundamental
observer, with light  rays  traveling  at  $45^o$.  Given  this
picture, it is simple to see  that the differing travel times noted by
\citet{abram} are simply an issue of synchronicity.

While the  two panels in Figure~\ref{fig1} both  represent a fully conformal
representation of an $\Omega<1$ universe, each display differing lines
of  simultaneity; in the  left-hand panel  the dashed  line represents
constant   $\tau$,  or   proper  time   as  measured   by  fundamental
observers. In standard FLRW universes, these hyperbolae represent times
(slices) of equal matter/energy density as seen by comoving observers. 
In  the right-hand panel, the lines of
simultaneity are represented by  lines of constant comoving coordinate
$t$.   An   examination  of  the  left-hand  panel,   whose  lines  of
synchronicity represent constant cosmological time in the FLRW metric,
reveals  that both  observers agree  that the  duration of  the outward
light  ray (red paths  of the  observers) is  shorter than  the return
journey (shown  in blue). Both observers  agree on the  length of time
each  leg  of  the  journey  took.  However,  an  examination  of  the
right-hand panel, where lines of synchronicity are defined by slices of
constant   conformal   coordinate  time   $t$,   reveal  a   different
picture. Now,  both observers agree  that the duration of  the outward
and return journey are equal, but  they disagree on how much time the 
journey took in total.

How are we to interpret this picture? Clearly the issue lies with the 
fact that measuring the journey time for any individual leg 
depends upon the comparison of clocks in differing inertial frames, 
a message stated by \citet{2006astro.ph..1171C} who considered a 
Milne (special relativistic universe) and inertial observers. This paper
shows that this is generally true in the relativistic interpretation of 
any FLRW universe and the ``{\it different} spacetime structure(s)"
(Figures 3-4 of \citet{abram}) are purely due to the way they have chosen define
synchronicity and slice up spacetime.  

\section{Carrying a Clock}\label{clock} 
Of course, a major problem with considering the path of a photon is that it
is null and the affine parameter that describes its path has no physical
significance.  But what if the photon is replaced with an observer who can
tick off their own proper time on the journey? To this end, we consider an
observer who accelerates away from the origin in a rocket with a constant
proper acceleration.  The acceleration is continued for a fixed amount of
proper time $\Delta \tau_r$ (for the rocketeer), followed by a coasting
period where the rocket is turned off. The rocket is then swung round as
the rocketeer accelerates back towards the origin, with the rocket is
fired for a time $2\Delta\tau_r$ before again entering a coasting period.
The rocket is turned round again so that the acceleration is away the
origin and fired for a time $\Delta\tau_r$. Within the flat space-time 
of special relativity, such a symmetric path will bring the rocketeer to
rest at the origin at the conclusion of their journey.

A general discussion of the influence of expanding space on the motion of
an accelerating observer will be presented in a future contribution (Kwan
\& Lewis {\it in preparation}), but here we consider three specific cases.
The top row of Figure~\ref{fig2} presents the journey of two rocketeers in
an expanding, open universe; in this case, $\Omega_o=0.001$ and the
resultant space-time structure should be akin to the Milne (empty)
universe. The first rocketeer leaves from the origin, and carries out the
symmetric accelerations outlined above. The second undertakes the same
journey, starting at the same cosmic time, $t'$, but at a different
comoving spatial location. To emphasise the issue of synchronicity when
making comparisons between different coordinate systems, both journeys
have been represented twice, in partially conformal coordinates on the left
and in fully conformal coordinates on the right. As expected, in either
representation the rocketeers return to rest at the the origin of their
journey. However, an examination of the rocketeer's path in the partial
conformal coordinates reveals that it is not symmetric, with more time $\eta$
spent reaching the most distant point from the origin, than the return journey.
Furthermore, the mid-point of 
the journey as seen by the rocketeer (where the path colour switches from red to blue) 
also does not correspond to the most distant point reached in the journey.

Moving to the fully conformal picture (right hand panel) we would expect
this path to be virtually the same as that seen in special relativity
\citep[see][for a discussion of the behaviour of the conformal
transformation for an open universe as
$\Omega_o\rightarrow0$]{2006astro.ph..1171C}; this is precisely what is
seen, with the path of the rocketeer starting at the origin being
symmetric in the conformal time $t$. Remember that in this representation,
observers at fixed comoving distances are now seen on sloping lines and
the rocket still reaches its greatest comoving distance during its
deceleration (dark blue), although the rocketeer reaches the maximum coordinate distance $r$
at the mid-point of the journey. The path of the rocketeer who starts at the
non-zero comoving coordinate is a little more complex, and quite different to that
seen in the partial conformal coordinates, but it also returns to
its origin after a symmetric flight.

The second row of Figure~\ref{fig2} presents identical journeys in an
open, matter dominated universe with $\Omega_o=0.500$, again with the left
hand panel presenting the partial conformal coordinates and the right hand
panel presenting the fully conformal representation.  While several aspects of the
paths of the rocketeer are similar to those seen in Figure~\ref{fig2},
there is a very important difference, namely that even though the paths
are symmetric in terms of acceleration and coasting time for the
rocketeer, they do not come to rest at the origin of their journey. In
fact, in this open case, the rocketeer over shoots and even when their
rocket is turned off, they are still moving away from their origin.
Exactly the same behaviour is seen in the fully conformal picture.  
Although, it may seem that this asymmetry, absent in the case of a static
universe (see Figure~\ref{fig1}), is an indication that space is really
expanding, this effect only occurs with the introduction of matter content
to act on the motion of the rocketeer.  In any case, we might naively
expect that if space is expanding, then the journey is longer on the way
back than it was on the way forwards [c.f. figure 4 of \cite{abram}] and
hence we might predict under-shooting, rather than over-shooting, the
origin. It must be emphasised that this over shoot is no different in the
Newtonian limit of the FLRW metric without expansion;  using Gauss's law
we can see that at a given radius $R$ from the origin, the rocketeer
experiences a gravitational acceleration towards the origin due to the
mass contained by a sphere of radius $R$.  This imaginary sphere changes
size throughout the journey but the acceleration from gravity remains
pointed towards the origin during the entire trip. Thus in addition to
the thrust provided by the rocket, the rocketeer recieves at all times an
additional push towards the origin.  We should not, therefore, be surprised to
find the rocketeer overshoots. Thinking very simply about the effects of
gravity, rather than the more nebulous expansion of space, gives a much
simpler intuitive view.

This asymmetry is even more apparent in the bottom row of
Figure~\ref{fig2} which again presents the rocketeers' paths in partial
and fully conformal coordinates, except now the matter density is
$\Omega_o=0.999$; in this case, the universe is approaching the spatially
flat $\Omega_o=1$ universe and hence the slope of the comoving observer in
the fully conformal picture is approaching that of the partially conformal
case (as noted previously, for spatially flat models the partial and fully conformal
transforms are equal). The key difference between the three cases
represented in Figure~\ref{fig2} is that the matter content of each
universe increases, which we are free to interpret as the cause of the
increasing asymmetry in the paths, rather than the asymmetry being 
caused by the expansion of space.  
Again this overshoot can be understood in the Newtonian limit of the FLRW
metric without expansion; we would expect approximately the same behaviour
to occur for a rocketeer travelling at non-relativistic speeds in a
Newtonian potential to a destination close by. The implications of this 
journey on the question of whether space really expands are presented in 
Section~\ref{syn}.

\subsection{Synchronizing Clocks}\label{sync}
It is clear that the timing issues related to the radar experiment are
related to the synchronization of clocks (as are many of the
problems and apparent paradoxes in relativity). However, the universe itself
provides a  clock that can be  employed to provide at  least a working
definition of synchronicity using the density of matter/energy; as the
universe expands, the density of  matter falls and dashed lines in the
right hand  panel of Figure~\ref{fig1} correspond to  slices of cosmic
time in the FLRW metric along which the density is equal.
Clearly, such  synchronization is not possible in  the Milne universe,
as, being empty, there is no  density yardstick with which to tick off
cosmic  time.   However, the  situation  is  the  same in  a  universe
containing only  a cosmological constant term (with  equation of state
$w=-1$) as the energy density remains a constant. 

With  either of  these  cases, or  any  other universe  model, we  can
imagine a  `test CMB', a homogeneous  and isotropic bath  of photons of
negligible  energy  fraction  defining   the  Hubble  rest  frame  and
comoving coordinates. By measuring the temperature of  the CMB all
observers  can calibrate  their clocks  to each  other. Interestingly,
even in the Milne universe which contains no  gravitating energy, this
test CMB  provides a universal clock giving  an excellent demonstration
of how  we can observe  an apparent expansion  of space in  a universe
known   to   be  completely   empty   and   equivalent   to   special
relativity. Clearly the interpretation of expanding space `stretching'
photons causing them  to redshift, while being a  useful teaching aid,
does not describe a casual  physical phenomenon if we can observe this
effect in Minkowski space.

\section{Conclusions: So, is Space Really Expanding?}\label{syn} This 
work has grown out of a recent exchange in the literature on the 
question on whether space `really' expands.  In a previous contribution 
\citep{2007PASA...24...95F}, we argued that while space is `completely 
and utterly empty' (to quote Steve Weinberg), it is perfectly valid to 
interpret the equations of relativity in terms of an expanding space.  
The mistake is to push analogies too far and imbue space with physical 
properties that are not consistent with the equations of relativity.

In their recent  work, \citet{abram} showed that, in  all but an empty
universe, distances derived from  the Hubble law and radar ranging
differ and hence ``one must conclude that space is expanding". But how
is this difference  occurring?  Is the expansion of  space acting on a
light ray (or  even a rocketeer) as they  travel through the universe?
We can  think of space as  a rubber sheet that stretches to wash
out peculiar motions  and drives everything back into  the Hubble flow
[see  \citet{barnes2006}].  
However, it is the presence of matter that necessitates the inclusion of 
gravitational forces upon the  motion  of  the  rocketeers   and  it  is  
this  -  the  changing gravitational influence of matter in the universe 
on the rocketeers  - that  causes the increasing  asymmetry moving  down the
panels in Figure~\ref{fig2}, not that space physically expands.

In closing,  we state that it is  a fools  errand to search  for the
truth  of the existance  of expanding  space; not  only because  it is
dependant  upon a  choice  of coordinates,  but  also because  general
relativity is represented by Newtonian physics in the weak field limit
and  the  global  behaviour  of  the FLRW  metric  always  reduces  to
Newtonian gravity in the limit of  the local universe  with no need
for  expanding space.  While the  expansion of  space is  a  valid (but
dangerous picture  when working with the equations  of relativity, any
attempts)  to obtain observations  to address  the question  of whether
galaxies are  moving through static space  or are carried  away by the
expansion of space are doomed to failure.

\section*{Acknowledgments} 
GFL acknowledges support from ARC Discovery Project DP0665574. MJF is
supported by a scholarship partially funded by the Science Faculty at The
University of Sydney. JK is supported by an Australian Postgraduate Award.

\end{document}